\documentclass[12pt]{article}
\usepackage[T1]{fontenc}
\usepackage{comment}
\usepackage[utf8]{inputenc}
\usepackage[superscript,biblabel]{cite}
\usepackage{physics}
\usepackage{multicol}
\usepackage[english]{babel}
\usepackage{amsthm}
\usepackage{graphicx}
\usepackage{frontespizio}
\usepackage{xcolor}
\usepackage[most]{tcolorbox}
\usepackage[]{amssymb}
\usepackage[]{enumerate}
\usepackage[]{tikz}
\usepackage[]{circuitikz}
\usepackage[]{float}
\usepackage{textcomp}
\usepackage{pdfpages}
\usepackage{fancyhdr}
\usepackage{blindtext}
\usepackage[]{amsfonts}
\usepackage{amsmath}
\usepackage{subfig}
\usepackage{comment}
\usepackage{authblk}
\usepackage{soul}
\usepackage{cite}
\usepackage[english]{babel}

\usepackage[letterpaper,top=2cm,bottom=2cm,left=1.75cm,right=2.5cm,marginparwidth=2.5cm]{geometry}

\usepackage{amsmath}
\usepackage{graphicx}
\usepackage[colorlinks=true, allcolors=blue]{hyperref}
\newcommand{\ed}[1]{\textcolor{black}{#1}}
\usepackage[left]{lineno}
\begin{document}
%

\title{ Paramagnetically driven superconducting re-entrance in Eu-doped infinite layer nickelates}
\vspace{5pt}
\date{\vspace{-5ex}}

\renewcommand\Authfont{\fontsize{12}{14.4}\selectfont}
\renewcommand\Affilfont{\fontsize{9}{10.8}\itshape}

\author[1]{Lucia Varbaro}

\author[1]{Lukas Korosec}
\author[1,2]{Chih-Ying Hsu}
\author[2]{Duncan T.L. Alexander}
\author[2]{Pau Torruella}
\author[1]{Clémentine Thibault}
\author[3]{Benjamin A. Piot}
\author[3]{David Le Boeuf}
\author[4]{Javier Herrero Martin}
\author[4]{Weibin Li}
\author[5]{Evgenios Stylianidis}
\author[5]{Marta Gibert}
\author[6]{Marc Gabay}
\author[1]{Jean-Marc Triscone}

\affil[1]{DQMP, University of Geneva, Geneva, Switzerland}
\affil[2]{Electron Spectrometry and Microscopy Laboratory (LSME), Institute of Physics (IPHYS), Ecole Polytechnique Fédérale de Lausanne (EPFL), Lausanne, Switzerland}
\affil[3]{Laboratoire National des Champs Magnétiques Intenses (LNCMI-EMFL), CNRS, Université Grenoble Alpes,
UPS, INSA, 38042 Grenoble, France}
\affil[4]{ALBA Synchrotron Light Source, Cerdanyola del Vallès 08290, Spain}
\affil[5]{Institute of Solid State Physics, TU Wien, 1040 Vienna, Austria}
\affil[6]{Laboratoire de Physique des Solides, Université Paris-Saclay, CNRS UMR 8502, Orsay, France}

\newcommand{\NNO}{NdNiO$_3$}
\newcommand{\iNNO}{NdNiO$_2$}
\newcommand{\NENO}{Nd$_{1-x}$Eu$_{x}$NiO$_2$}
\newcommand{\NENOp}{Nd$_{1-x}$Eu$_{x}$NiO$_3$}
\newcommand{\SNENO}{Sm$_{1-2x}$Nd$_{x}$Eu$_{x}$NiO$_2$}
\newcommand{\oNENOp}{Nd$_{0.7}$Eu$_{0.3}$NiO$_3$}
\newcommand{\oNENO}{Nd$_{0.7}$Eu$_{0.3}$NiO$_2$}
\newcommand{\SNO}{SmNiO$_3$}
\newcommand{\RNO}{$R$NiO$_3$}
\newcommand{\ABO}{ABO$_3$}
\newcommand{\NSNO}{Nd$_{1-x}$Sm$_x$NiO$_3$}
\newcommand{\NLNO}{Nd$_{1-x}$La$_x$NiO$_3$}
\newcommand{\LAO}{LaAlO$_3$}
\newcommand{\STO}{SrTiO$_3$}
\newcommand{\NGO}{NdGaO$_3$}
\newcommand{\LNO}{LaNiO$_3$}
\newcommand{\TMI}{T$_{MI}$}
\newcommand{\deltaN}{$\Delta_N$}

\maketitle
\textbf{The breakthrough discovery of superconductivity in infinite-layer nickelates, and subsequently in several superconducting nickelates with more complex layered structures, capped a search spanning more than two decades and opened an entirely new field of research. Significant efforts aim to increase the critical temperature, to determine the electronic structure of the system, the underlying pairing mechanism, and the similarities between this system and cuprates $-$ Ni$^{1+}$ in infinite-layer nickelates being isoelectronic to Cu$^{2+}$ in high-T$_c$ cuprates. Here, we \ed{explore} the unique role of magnetic rare earth ions in superconducting Eu-doped NdNiO$_2$. We show that the field-induced re-entrant superconductivity which we evidence in this compound is the result of a delicate balance between the competing effects of the Eu$^{2+}$ and Nd$^{3+}$ ions. Our analyses of the extraordinary Hall effect and modeling of the superconducting critical fields demonstrate that the influence of these ions on magneto-transport is only felt when they are polarized by a magnetic field.}

\vspace{5mm}

\begin{multicols}{2}
The 2019 surprise \ed{report} of superconductivity in Sr-doped infinite layer \iNNO{} thin films drew worldwide attention \cite{li_superconductivity_2019}. Soon after the \ed{discovery} of Danfeng Li et al. at Stanford, several groups reproduced the results and extended the scope of superconducting infinite layer nickelate materials \cite{li_superconducting_2020, wang_effects_2023, lee_aspects_2020, sun_situ_2024, zeng_superconductivity_2022, osada_phase_2020, osada_nickelate_2021, wang_pressure-induced_2022, zeng_phase_2020, lee_linear--temperature_2023,zhang_achieving_2024,wei_solid_2023,wei_superconducting_2023}. Very recently, Ariando’s group in Singapore announced a critical temperature T$_c$ of up to 40 K in an infinite layer nickelate material containing, Sm, Eu, Sr and Ca  in addition to Ni and O  \cite{chow_bulk_2025}, a result partially reproduced \cite{yang_enhanced_2025}. Also, several other layered nickelates compounds have been found to be superconducting both in bulk and thin film form, enriching and broadening this new field of research (see, for instance, [\citen{pan_superconductivity_2022,liu_superconductivity_2025,zhu_superconductivity_2024,zhang_high-temperature_2024}]). 
    
Alongside the pursuit of higher critical temperatures, significant experimental and theoretical efforts have been devoted to \ed{unraveling} the fundamental properties of these new superconductors and the underlying mechanism of superconductivity.\\ Experimental studies of infinite layer compounds have focused on the effects of Sr doping, \ed{on} the phase diagram, the characterization of superconducting properties, the investigation of charge ordering phenomena, as well as \ed{on the study of} magnetic correlations in the undoped parent compound. \cite{hepting_electronic_2020, ding_cuprate-like_2024, goodge_doping_2021, dong_topochemical_2025, sun_electronic_2025, lu_magnetic_2021, fowlie_intrinsic_2022,li_topotactic_2024,parzyck_absence_2024,rossi_broken_2022,tam_charge_2022}.\\ On the theory side, several studies focused on the electronic structure of the system, the origin of superconductivity, and the commonalities and differences with cuprates (Ni$^{1+}$ in infinite layer nickelates being isoelectronic to Cu$^{2+}$ in the high-T$_c$ superconducting cuprates). \ed{Understanding the role played by the much larger charge transfer energy in nickelates pushing the O p-band way below the Fermi level and the hybridization between the Ni 3d and rare earth states are issues of definite interest} (see, for instance, [\citen{botana_similarities_2020,lee_infinite-layer_2004,foyevtsova_distinct_2023,li_two-gap_2024, zhang_self-doped_2020, jiang_electronic_2019, bandyopadhyay_superconductivity_2020, wu_robust_2020,jiang_critical_2020}]).

In this report we explore the magnetotransport properties of infinite layer Nd-nickelate films doped with Eu (NENO) and the important role played by the rare earth magnetic ions, \ed{following the route proposed at Yale by \textit{Wei et al.} in [\citen{wei_superconducting_2023}] for the doping and the reduction process.}  We observe in these Eu-doped compounds a very unusual magnetic field driven re-entrant superconductivity - \ed{as was} reported at the 2023 MRS Fall meeting \cite{ahn_synchrotron_nodate} and in recent preprints [\citen{yang_robust_2025,vu_unconventional_2025,rubi_extreme_2025}]. More \ed{specifically}, although superconductivity occurs in zero magnetic field in the presence of magnetic Eu$^{2+}$ and Nd$^{3+}$ (30\% and 70\% respectively, in optimally doped samples), \ed{applying} a weak magnetic field polarizes the rare-earth ions and weakens the superconducting state which, however, reappears at higher fields. Such behavior appears to be specific to “low-T$_c$” Eu-doped films,  and is linked to the so-called Jaccarino-Peter effect. \ed{These authors originally proposed a scenario whereby a rare-earth ferromagnetic metal with negative exchange interaction subjected to a magnetic field may turn into a superconductor \cite{jaccarino_ultra-high-field_1962}. } However, unlike the original Jaccarino-Peter proposal and the experimental observation of magnetic field induced superconductivity in Chevrel phases \cite{jaccarino_ultra-high-field_1962,meul_observation_1984}, our extraordinary Hall effect data, critical field measurements and modeling demonstrate that this unusual behavior stems from a delicate balance between the \ed{opposing} influence of Eu$^{2+}$ and Nd$^{3+}$ and a particular compensation mechanism linked to the presence and magnetic response of both types of magnetic ions. This phenomenology emerges only in samples with relatively low T$_c$, where the temperature remains sufficiently low to allow for a strong paramagnetic response of the rare-earth moments.\\

To explore the interplay between magnetism and superconductivity in infinite-layer nickelates, we investigated mainly optimally doped Nd$_{1-x}$Eu$_x$NiO$_2$ ($x = 0.3$) thin films grown on two distinct substrates: \ed{LSAT ($\left(\mathrm{La}_{0.3}\mathrm{Sr}_{0.7}\right)\left(\mathrm{Al}_{0.65}\mathrm{Ta}_{0.35}\right)\mathrm{O}_{3}$
) and NdGaO$_3$. In the former case, our films are superconducting while in the later case they are not, the exact nature of this difference in behavior still remains to be understood}. Nevertheless, samples grown on NdGaO$_3$ turned out to be well-suited for magnetotransport (Hall) studies at low temperatures. This complementary \ed{approach enabled us to perform } a detailed study of the superconducting properties and of the paramagnetic response of the Nd$^{3+}$ and Eu$^{2+}$ ions. \\The optimization of the superconducting phase was initially carried out on LSAT substrates. In this approach, the oxidation of a metallic aluminum layer is used to remove the apical oxygens present in the perovskite phase, resulting in films with high crystalline quality. In our case, 20 to 28-unit-cell-thick \oNENOp{} films were deposited on (001) LSAT  and (110)$_{orthorhombic}$ NdGaO$_3$ via off-axis RF magnetron sputtering. To deposit metallic Al and enable the subsequent topotactic reduction leading to the infinite-layer phase, an additional sputtering gun, in an on-axis geometry, was used, facing directly the substrate holder. More information about the Al growth can be found in the Methods section.\\

\begin{figure}[H]
\centering
\includegraphics[width=1\textwidth]{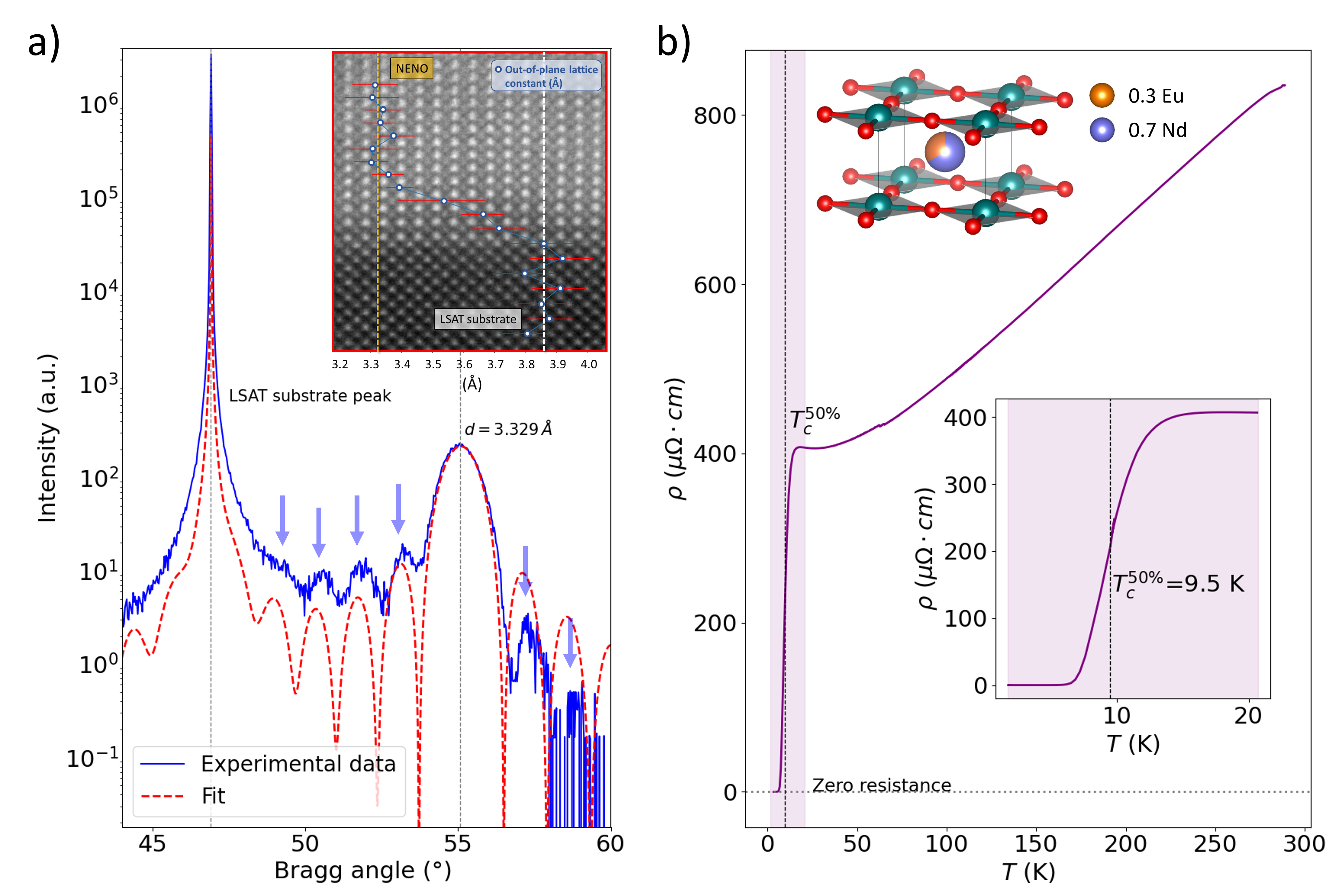}
\caption{(a) High-resolution X-ray diffraction scan around the (002) reflection of an optimally doped Nd$_{0.7}$Eu$_{0.3}$NiO$_2$ film grown on LSAT. The scan reveals  an intense film peak, and the pronounced finite-size oscillations, marked by arrows, indicate high crystalline quality and smooth interfaces. The fit (red dashed line) yields an out-of-plane lattice parameter of 3.329 Å and an estimated film thickness of approximately 22 unit cells. The determined c-axis value is  confirmed by fitting the atomic positions and measuring the out-of-plane lattice spacing in a high magnification STEM image shown in the inset. (b) Resistivity versus temperature measurements for an optimized Nd$_{0.7}$Eu$_{0.3}$NiO$_2$ film on LSAT showing a sharp superconducting transition with  T$_c^{\%50 }$ = 9.5 K and a zero-resistance state below $\sim$ 6 K.}
\label{fig:1}
\end{figure}

Figure~\ref{fig:1}(a) presents a high-resolution X-ray diffraction (XRD) scan around the (002) reflection of a representative Nd$_{0.7}$Eu$_{0.3}$NiO$_2$ film on LSAT. The scan shows an intense (002) film peak and the presence of well-defined finite-size fringes over a wide angular range are indicative of excellent structural quality and a sharp film-substrate interface. Fitting the diffraction pattern [\citen{lichtensteiger_interactivexrdfit_2018}] (red dashed line) $-$ yields an out-of-plane lattice constant of 3.329 Å and a film thickness of approximately 22 unit cells. This is  confirmed by a scanning transmission electron microscopy (STEM) study (see inset Figure \ref{fig:1}), where the atomic positions of the A sites have been fitted to extract the spacing between the atomic planes \cite{noauthor_httpsatomaporg_nodate}. Further STEM characterization of samples on LSAT both for the perovskite and infinite layer phases can be found in the Supplementary Information section (Fig S1 and S2). \\Figure \ref{fig:1}(b) shows the resistivity versus temperature curve of an optimally doped, reduced sample. In the top left corner, a schematic of the infinite layer crystal structure is shown and the selected stoichiometry indicated. The resistivity versus temperature curve shows a complete superconducting transition, with a midpoint critical temperature T$_c^{50\%}$ of 9.5 K and a zero-resistance state reached below approximately 6 K, as highlighted in the inset. Transport and X-ray diffraction data for an optimally doped \NENO{} sample grown on \NGO{} can be found in Figure S3 of the Supplementary Information section.

\end{multicols}

\begin{multicols}{2}

\begin{figure}[H]
\centering
\includegraphics[width=0.9\textwidth]{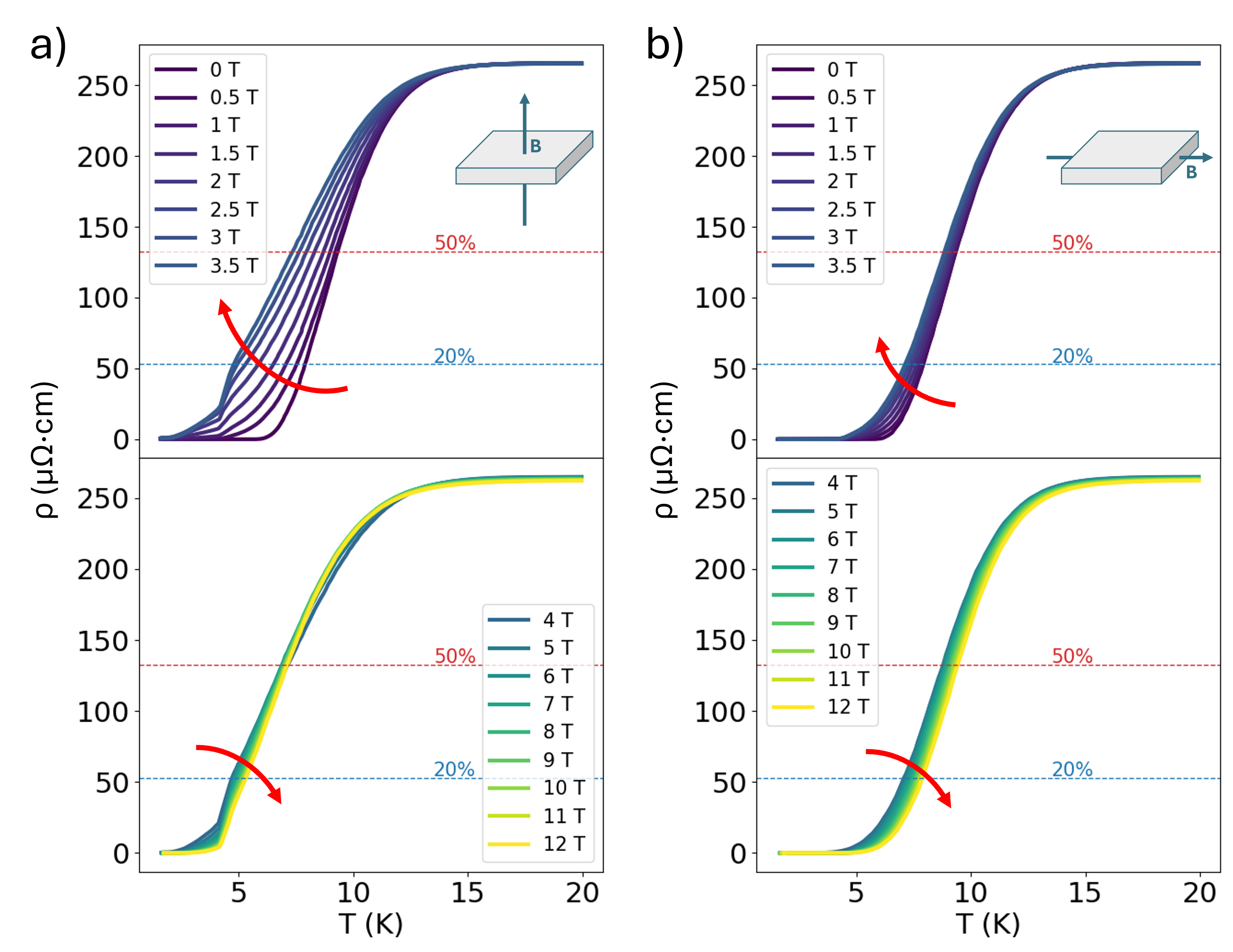}
\caption{(a) Resistivity versus temperature curves for an optimized Nd$_{0.7}$Eu$_{0.3}$NiO$_2$ film on LSAT under perpendicular magnetic fields ranging from 0 T to 3.5 T in the top panel and 4 to 12 T in the bottom one. A conventional suppression of the superconducting transition temperature $T_c$ is observed in the lower field range, followed by a re-entrant enhancement of superconductivity at higher fields. (b) Analogous measurement performed in the parallel magnetic field configuration, displaying the same re-entrance at around 3.5 T. \ed{This sample is from a different batch than the one} whose properties are shown in Fig. \ref{fig:1}b).}  

\label{fig:2}
\end{figure}

To further investigate the robustness of superconductivity in the optimized, optimally doped Nd$_{0.7}$Eu$_{0.3}$NiO$_2$ films, we measured the resistivity as a function of temperature on two different samples, under both perpendicular and parallel magnetic fields ranging from 0 T to 12 T (in Geneva - UNIGE (sample 1)), and up to 30 T at the LNCMI-EMFL high magnetic field facility in Grenoble (sample 2). Figure~\ref{fig:2}(a) displays resistivity curves acquired on sample 1 between 2 K and 20 K in the out-of-plane configuration. At low fields (top panel, 0–3.5 T), the superconducting transition progressively shifts to lower temperatures and broadens with increasing field, as expected. However, above 3.5 T, a non-monotonic behavior emerges: further ramping up of the field stabilizes and sharpens the superconducting transition, \ed{pushing} it back to higher temperatures. This unconventional field-induced enhancement is the signature of a re-entrant superconducting phase at intermediate fields. \\Figure \ref{fig:2}(b) presents a rather similar behavior for the in-plane configuration, although the field appears to be less effective in suppressing superconductivity as expected for the orbital channel and observed in  measurements of thin superconducting films in parallel field configuration (see for instance [\citen{reyren_superconducting_2007, wang_isotropic_2021}]). \ed{Note} that all of our resistivity curves display a small anomaly at around 3 K. This may be caused by some magnetic ordering of rare-earth magnetic moments, which takes place at temperatures between 1 $-$ 6 K in several complex oxides of Nd$^{3+}$ and Eu$^{2+}$. \cite{bartolome_spin12_1994,plaza_neutron_1997,mcguire_magnetic_1966,zong_antiferromagnetism_2010}.





\begin{figure}[H]
\centering
\includegraphics[width=1\textwidth]{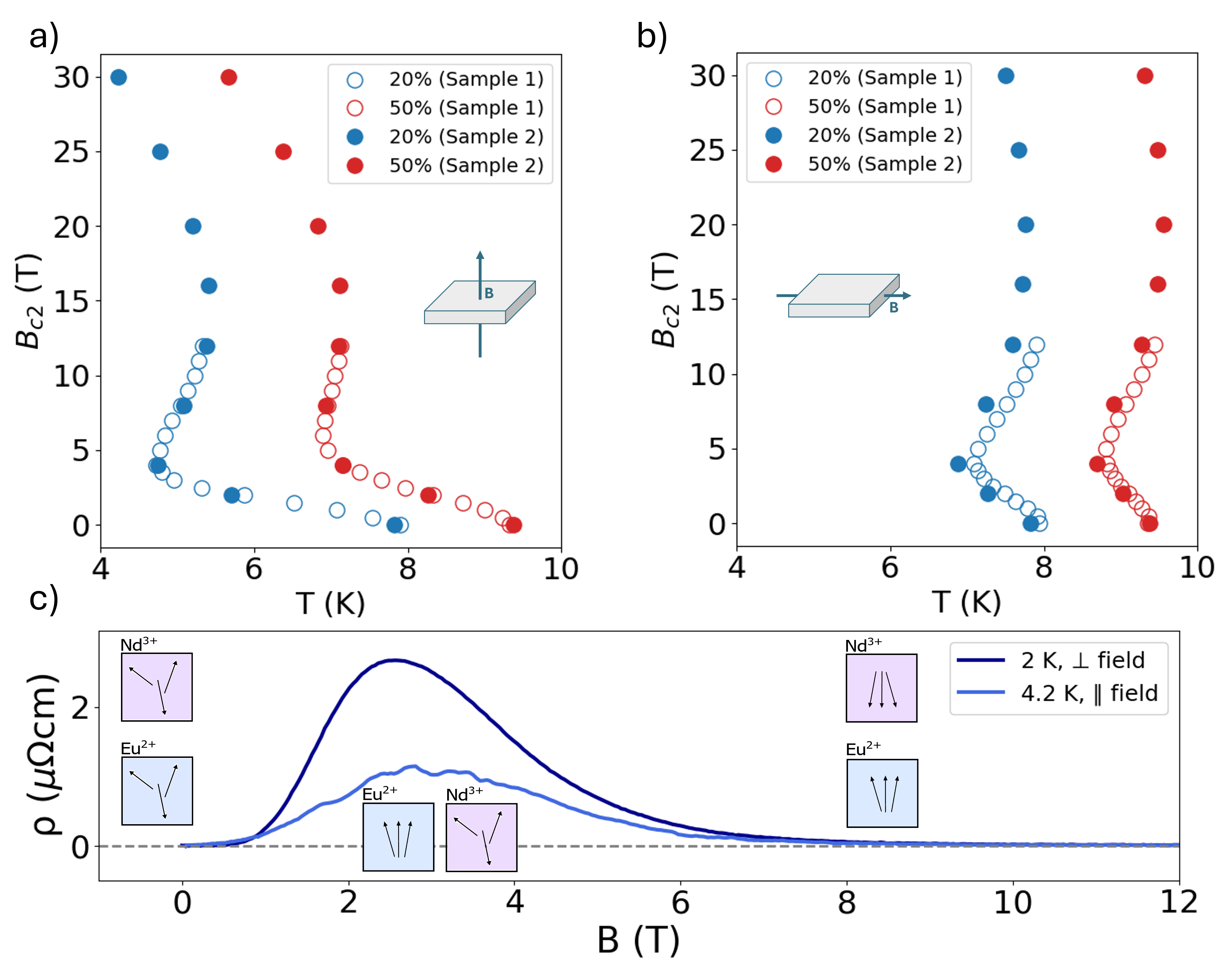}
\caption{(a) and (b): Magnetic phase diagrams for out-of-plane (perpendicular) (a) and in-plane (parallel) (b) magnetic field configurations. Blue and red markers correspond respectively to the superconducting transition temperatures extracted using the 20\% and 50\% criteria (temperatures at which the resistivity is 20\% and 50\% of its value in the normal state at 15 K, as indicated by the dashed lines in Figure \ref{fig:2}). For both criteria, the solid markers correspond to measurements performed at the LNCMI-EMFL high-magnetic-field facility in Grenoble, while the open markers refer to measurements carried out at UNIGE on a different sample with the same composition. In both configurations, a low-field re-entrant superconducting behavior is observed, suggesting the presence of two superconducting regions, the second persisting up to the measured field of 30 T, separated (in some temperature range) by an intermediate dissipative phase. As can be seen and as expected, the field is less effective in suppressing superconductivity in the in-plane geometry.  
(c): Magnetoresistivity curves at fixed temperatures of 2 K and 4.2 K, for perpendicular and parallel geometries, respectively. Together with the data, a schematic physical picture representing the rare-earth spins configuration is presented. The light blue and purple insets illustrate respectively the Eu$^{2+}$ and Nd$^{3+}$ spin polarizations in each regime and their compensation as the external field is increased.}
\label{fig:3}
\end{figure}

To visualize the evolution of this re-entrant behavior, we extracted the superconducting transition temperatures ($T_c$) using two different criteria (20\% in blue and 50\% in red) of the normal-state resistivity at 15 K, indicated by the horizontal dashed lines in Figure \ref{fig:2}, and constructed the corresponding magnetic phase diagrams shown in Figure~\ref{fig:3}(a) and (b). For both configurations, at low temperatures and intermediate fields, field sweeps at fixed temperature reveal two superconducting regions separated by a dissipative regime. This behavior is illustrated in the low-temperature magnetotransport data shown in Fig.~\ref{fig:3}(c): the system first enters a dissipative state at around 3 T, and then re-enters a superconducting state that persists up to 30 T (shown only up to 12 T in the Figure).\\ The high-field data taken up to 30 T at the LNCMI-EMFL high-magnetic-field facility in Grenoble on a second film with the same Eu content ($x_{\mathrm{Eu}} = 0.3$) are also included on Figure~\ref{fig:3}(a) and (b). The excellent agreement between the LNCMI-EMFL data (solid markers) and the UNIGE data (open markers) for both criteria demonstrates the reproducibility of the samples and their phase diagram and confirms that the low-temperature superconducting state persists at least up to 30 T. The small difference observed in parallel field might be due to a slight misalignment between the field direction and the film plane. \\As previously mentioned, such re-entrant superconductivity has been observed in other compounds containing Eu elements, most notably in Chevrel phases \cite{meul_observation_1984}, where magnetic interactions involving Eu$^{2+}$ moments are believed to modulate the superconducting state under applied fields. Key to this effect is the negative exchange interaction between the Eu$^{2+}$ moments and the spin of the conduction electrons (also at the origin of Kondo physics) and the polarization in magnetic field of the rare-earth ions \cite{buschow_intermetallic_1977,finley_spintronics_2020,hansen_chapter_1991,sala_ferrimagnetic_2022,kim_ferrimagnetic_2022},  that can lead to a reduction of the total field experienced by the conduction electrons - the Jaccarino-Peter effect \cite{jaccarino_ultra-high-field_1962}.  
\end{multicols}

\begin{multicols}{2}
To complement the main transport data, we performed voltage–current ($V$–$I$) measurements as a function of temperature and magnetic field (see Figure S4). At zero field and low temperatures, the $V$–$I$ curves show a wide zero-voltage plateau, consistent with robust superconductivity. Increasing the field to 2 T strongly suppresses the critical current and narrows the plateau, \ed{signaling} a weakening of the superconducting phase. We note that at higher fields (12 T), \ed{a recovery of} the critical current is observed, in line with the re-entrant behavior of superconductivity that we described in the above paragraph, as best seen in Figure S4(f) and (g), respectively at 1.7 and 2.5 K. The critical current density $J_c$ at 1.7 K for 0 T and 12 T is 1-2 10$^5$ A/cm$^2$, a very large value. The similarity of the critical current densities for these two magnetic fields suggests similar bulk-like superconducting states. \\

We now turn to the role of the rare-earth magnetic ions, Eu$^{2+}$ and Nd$^{3+}$, on the transport properties and on superconductivity, bearing in mind the fact that even for 30\% of Eu$^{2+}$ and 70\% of Nd$^{3+}$ one observes superconductivity in zero field. To understand the nature of charge transport and the role played by the rare-earth ions in optimally doped Nd$_{0.7}$Eu$_{0.3}$NiO$_2$, we performed Hall effect measurements on both non-superconducting and superconducting thin films grown on NdGaO$_3$ and LSAT substrates, respectively. If the observation of re-entrant superconductivity is linked to a Jaccarino-Peter like effect, the re-entrant superconducting behavior should be \ed{driven by} a polarization of the magnetic moments $-$ with the complexity here that the system contains \ed{two species of } magnetic ions, Eu$^{2+}$ and Nd$^{3+}$. \\

\begin{figure}[H]
\centering
\includegraphics[width=1\textwidth]{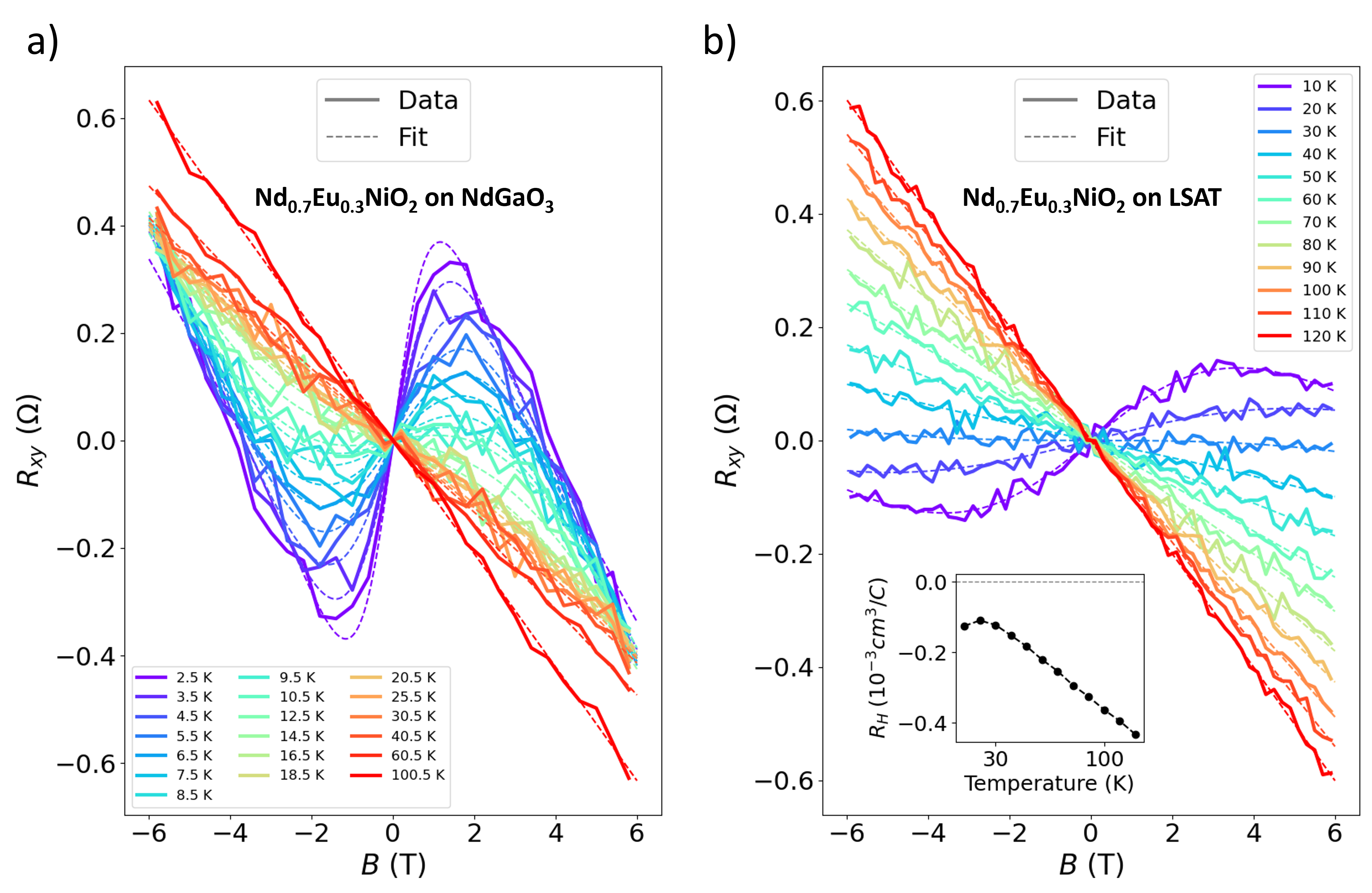}
\caption{Hall resistance R$_{xy}$(B) measured as a function of out-of-plane magnetic field in optimally doped Nd$_{0.7}$Eu$_{0.3}$NiO$_2$ thin films grown on different substrates.
(a) Data from a non-superconducting optimally doped film grown on NdGaO$_3$, measured between 2.5 K and 100.5 K. The curves show a pronounced, temperature-dependent non-linearity, ascribed to a magnetic (extraordinary) Hall component possibly arising from the field-polarized Nd$^{3+}$ and Eu$^{2+}$ moments. Dashed lines represent fits using a two-component magnetic model  described in the main text.
(b) Data from a superconducting optimally doped film grown on LSAT, measured between 10 K and 120 K. 
Inset: temperature dependence of the extracted Hall coefficient $R_H$, obtained from fits (dashed lines) of the antisymmetrized data.}
\label{fig:4}
\end{figure}

The Hall resistance can be expressed as $R_{xy} = R_0 B + R_s  M$, where $R_0$ is the ordinary Hall coefficient, $R_s$ the extraordinary Hall coefficient, and $M$ the magnetization. The second term captures the influence of a magnetic order or field-induced magnetization on the Hall response. \ed{Our game plan  is to extract information on the rare-earth ions polarization from the extraordinary part of the Hall response}. Since the Hall response is masked by superconductivity below T$_c$ in films on LSAT, \ed{we measure the low T Hall response on a non-superconducting sample of the same composition, grown on NdGaO$_3$.} The results are shown in Figure~\ref{fig:4}, where panel (a) displays the $R_{xy}(B)$ data for the non-superconducting sample grown on \NGO{}, and panel (b) for the superconducting sample grown on LSAT.\\ The data were anti-symmetrized to remove possible longitudinal resistance contributions arising from contact misalignment and isolate the Hall response.\\

The Hall resistance data of the non-superconducting sample displays pronounced field-dependent nonlinearities, particularly at low temperatures, an extraordinary Hall contribution, likely stemming from the Eu$^{2+}$ and  Nd$^{3+}$ magnetic moments. For the superconducting film grown on LSAT, the transverse resistance was measured in the normal state between 10 K and 120 K. As well as for the lower temperature curves measured on the sample grown on \NGO{}, the 10 K curve for the superconducting sample displays non-linear behavior. \\ 
\end{multicols}

\begin{multicols}{2}

To quantitatively interpret the non-linear Hall response, we developed a model that accounts for the magnetic contributions of the two rare-earth species, Eu$^{2+}$ and Nd$^{3+}$, both of which, as mentioned above, carry localized magnetic moments. \\ 
We note that $J=S=\frac{7}{2}$ for Eu$^{2+}$ and $J=L-S$ ($J=\frac{9}{2}$) for Nd$^{3+}$.\\
As originally discussed by de Gennes \cite{gennes_interactions_1962}, the interaction between localized rare-earth moments and conduction electrons with spin $s_e$ can be described by a Heisenberg-type exchange Hamiltonian, $\mathcal{H}_{\text{ex}} = -\ed{\mathbf{\Gamma}} \, \vec{s}_e \cdot \vec{S}_j$. Since in rare-earth ions the total angular momentum $\vec{J} = \vec{L} + \vec{S}$ is the relevant quantum number, the spin operator is projected onto $\vec{J}$, resulting in the substitution $\vec{S} \rightarrow (g - 1)\vec{J}$, $g$ being the Landé factor ($g_{Eu}$ and $g_{Nd}$ being 2 and 0.72 respectively). This projected spin $-$ denoted $\vec{S}_{\text{proj}} = (g - 1)\vec{J}$ $-$ governs the effective exchange interaction and is used in our model to capture the extraordinary Hall contribution.
We express the total Hall resistance as the sum of an ordinary, linear-in-field term and a magnetic contribution given by the weighted sum of the averaged projected spins of the two rare-earth "sublattices":

$$
\begin{aligned}
R_{xy}(B, T) &= R_0(T)\,B
+ c \Big[ x_{\mathrm{Nd}} \,\langle S_{\mathrm{Nd}^{3+}}(B, T) \rangle \\
&\qquad\qquad\;\; + x_{\mathrm{Eu}} \,\langle S_{\mathrm{Eu}^{2+}}(B, T) \rangle \Big]
\end{aligned}
$$
\\
where $R_0(T)$ is the temperature-dependent ordinary Hall coefficient and $c$ is a temperature independent coefficient setting the scale of the total magnetic contribution.\\

The spin expectation values are evaluated for Eu and Nd using the Brillouin function $B_J(x)$:

$$
\langle S \rangle = - (g - 1)J \cdot B_J(x), \quad x = \frac{J g \mu_B B}{k_B T},
$$

\begin{equation*}
\begin{aligned}
B_J(x) &= \left( \frac{2J + 1}{2J} \right)
\coth \left( \frac{2J + 1}{2J}\, x \right) \\
&\quad - \left( \frac{1}{2J} \right)
\coth \left( \frac{x}{2J} \right).
\end{aligned}
\end{equation*}
\\

When computing $R_{xy}(B, T)$, the Nd$^{3+}$ and Eu$^{2+}$ concentrations ($x_{Nd}$ and $x_{Eu}$) were set to 0.7 and 0.3, respectively, assuming fully divalent Eu ions.  The Eu valence state was investigated in Supplementary Information Fig S5, which presents STEM electron energy-loss spectroscopy measurements of infinite layer \oNENO{} compared to unreduced \oNENOp{} films on LSAT, in the standard core-loss (O K, Ni L$_{2,3}$, Eu M$_{4,5}$) and deep core-loss (Eu L$_3$) regimes. While the O K and Ni L$_{2,3}$ edges show the characteristic features of infinite layer and perovskite nickelates in the respective samples [\citen{goodge_doping_2021}], both the Eu M$_{4,5}$ and L$_3$ edges indicate that Eu exhibits a 3+ valence state in the fully oxidized compound, that reduces towards Eu$^{2+}$ in the infinite layer \oNENO{} film.\\ 
Dashed lines in both Figures~\ref{fig:4}(a) and (b) represent fits obtained using this model, showing excellent agreement across the full range of fields and temperatures. For each temperature, $R_0(T)$, reflecting the ordinary Hall response, is extracted. The $c$ value, associated with the magnetic (extraordinary) component, is obtained from the fit performed at the lowest temperature and then kept constant across the full temperature range. The quality of the fits can be best appreciated when we extract the maximum value of $R_{xy}(B)$ at each temperature. Figure S6 compares the maximum value of $R_{xy}(B)$ versus temperature for the data and the model. A convincing agreement can be observed.\\
Importantly, since the $c$ coefficient is temperature independent, the temperature dependence of the non-linearity observed in the $R_{xy}(B)$ curves  arises solely from the partial compensation between the Brillouin functions of the two magnetic ions. 

\ed{In order to highlight} the contribution of each magnetic ion, we  calculated the \ed{respective }spin polarization of Nd$^{3+}$ and Eu$^{2+}$ as a function of field at select temperatures using their Brillouin functions. This is shown in Figure S7 of the Supplementary Information section. As can be seen on the Figure, the \ed{paramagnetic contributions of the two ions species partially cancel upon increasing the applied field, leading to a gradual suppression of the effective internal field produced by the Nd$^{3+}$ and Eu$^{2+}$ ions.} \\ The spin polarization of the two magnetic species was directly measured by X-ray magnetic circular dichroism (XMCD) at the ALBA Synchrotron light source. Figure S8(a) and (b) of the Supplementary Information show the x-ray absorption spectrum (XAS) and XMCD signal of an infinite layer \oNENO{} at the nominal temperature of 2 K and 6 T around the Eu and Nd M edges, respectively. From the XMCD signal measured at different magnetic fields, the spin polarizations of Nd and Eu magnetic ions were extracted, and are compared to the calculated Brillouin functions (S8(c)) plotted for the same temperature. To account for the surface sensitivity of the total electron yield XAS, possibly overestimating the amount of Eu$^{3+}$, the measured Eu signal has been rescaled by a factor of 2.

It is worth noting that a  low-temperature non-linearity in the Hall effect was reported by Dong et al. in [\citen{dong_topochemical_2025}] in non-superconducting NdNiO$_2$ films grown on NdGaO$_3$. In their case, the sample contains only Nd$^{3+}$ magnetic moments, and the observed non-linearity is consistent in sign with the contribution from Nd$^{3+}$ in our model.\\ The model can also be applied to samples with different compositions, as shown in Figure S9 which displays  Hall effect measurements  of a sample containing 35\% of Eu  and by the direct comparison of the transport properties in field between a sample containing 25\% of Eu and an optimally doped sample shown in Figure S10 of the Supplementary Information section. As expected from the sum of the weighted spin average values of Nd$^{3+}$ and Eu$^{2+}$ calculated through the Brillouin function formalism (Fig. S10(c)), when the amount of Eu is reduced, the transition to the first dissipative state appears at lower field values, as is clearly seen in the magnetoresistance data (Fig. S10(a)) and in the B$_{c2}$ vs T phase diagram panels (Fig. S10(b)).\\ Our model can be straightforwardly extended to include compounds featuring additional magnetic rare earths ions such as Sm, as shown in Figure S11 of the Supplementary Information section, along with experimental data on a superconducting Sm$_{0.3}$Nd$_{0.35}$Eu$_{0.35}$NiO$_{2}$ thin film.

Coming back to Figure \ref{fig:4}(b), we employed the model introduced above to fit the data and access the sign of the majority carriers. The non-linear behavior observed at the lowest temperature could be reproduced using values  for the parameters very similar to the ones which allowed us to to fit the Hall data of the non-superconducting sample in Figure~\ref{fig:4}(a). Figure S12 shows that the amplitude and the shape of the non-linear components of the Hall resistance for the superconducting and non-superconducting samples are very similar.\\
No sign reversal in the slope of $R_{xy}(B)$ is observed as a function of temperature, indicating that the dominant charge carrier type is always negative (electron-like) for this particular doping. \\

To link these observations with the measured superconducting behavior, one first notices that 30 T are not sufficient to drive the system back to the normal state. Rather, high field measurements performed on both Eu- and Sr-doped samples \cite{wang_effects_2023,wei_large_2023,wei_superconducting_2023, chow_pauli-limit_2022} confirm that H$_{c2}$ is very large, meaning that the orbital effect and Pauli limit do not affect superconductivity markedly in the field range investigated here. \ed{We thus ascribe} the observed behavior to the interactions between the rare-earth magnetic moments and the conduction electrons and to the compensation of the Eu$^{2+}$ and Nd$^{3+}$ contributions.\\ 

A plausible scenario explaining the collected experimental data is illustrated in the sketch of Figure \ref{fig:3}(c) together with a measurement of the low temperature longitudinal resistivity versus magnetic field (0$-$12 T) applied in both perpendicular and parallel directions for an optimally doped sample. At low fields, the magnetic moments are disordered and the material is superconducting. At slightly higher fields (around 3 T), as supported by both the Hall measurements and calculations, the Eu$^{2+}$ \ed{ions acquire a field-induced  polarization}. 
The Pair breaking terms consist of an orbital part plus a spin-orbit part. For that latter contribution, the magnetic field is the sum of the external and the exchange interaction fields (see below), driving the system into a dissipative state. At even higher fields, the Nd$^{3+}$ spins also "align", canceling the effect of the Eu$^{2+}$ ions, and restoring superconductivity. Finally, at much higher fields (> 30 T, not observed experimentally), the applied field should destroy the superconducting state. \\ Such a physical picture also naturally explains why re-entrant superconductivity is not observed in cases where T$_c$ is "too high" (see for instance Reference \cite{wei_superconducting_2023}). Indeed, as can be seen in the calculations of Figure S11(a), above 10 K, the resultant polarization of the magnetic moments is extremely weak. \\

To try quantifying the effects proposed above, we introduce a  microscopic description of the superconducting phase boundary in the $B_{c2}$–$T$ plane using Gorkov’s pair-breaking theory \cite{abrikosov_contribution_1961} for both the in- and out-of-plane field configurations. In this approach, superconductivity is suppressed by mechanisms that break Cooper pairs. Their effect is described in the BCS gap equation by a single pair-breaking parameter $\alpha$. \\

The modified BCS gap equation reads:
$$
  \ln\frac{T_c}{T_{c0}}
  = \psi\!\Bigl(\tfrac12\Bigr)
  - \psi\!\Bigl(\tfrac12 + \frac{\alpha}{2\pi\,k_B}T_c\Bigr)\,,
$$
where $T_{c0}$ is the zero field $T_c$ and $\psi(z) $ is the digamma function.\\ 

To capture both orbital ($\alpha_1$) and spin–orbit ($\alpha_2$) pair breaking\cite{tinkham_introduction_1975}, one writes for the perpendicular field configuration:
\begin{equation*}
\begin{aligned}
\alpha^{\perp} &= \alpha_1^{\perp} + \alpha_2, \\
\alpha_1^{\perp}(B_{0}) &= e\,D\,B_{0}, \\
\alpha_2(B_{\rm tot}) &= \frac{e^2\,h}{4\pi\,m^{*2}}\,\tau_{SO}\,B_{\rm tot}^2 .
\end{aligned}
\end{equation*}

and for the parallel field configuration:

$$
  \alpha^{\parallel} = \alpha_1^{\parallel} + \alpha_2,
  \qquad
  \alpha_1^{\parallel}(B_{\rm 0}) = \frac{D \left(e B_{0} d\right)^2}{6 h}
  \qquad
$$
\\
\ed{The expressions of $\alpha_1$ that we use apply to thin films when the mean free path $l$ is smaller than the thickness $d$ of the film. For our samples, $l\sim 1$ nm and $d\sim 7$ nm.}  
D is the diffusion constant, $B_0$ the applied external field, $B_{tot}$ the total magnetic field which affects the spin part of the wave function, $\tau_{SO}$ is the spin orbit diffusion time, and $m^*$ the electron mass. We determine the value of $D$ from the slope of the perpendicular upper critical field near $T_c$, using the dirty-limit relation $\left.\frac{dB_\perp}{dT}\right|_{T_c}=-\,0.40\,\pi k_B/(eD)$, which yields $D\simeq2\times10^{-5}\,\mathrm{m}^2/\mathrm{s}$ for our films. The spin–orbit scattering time was treated as an adjustable parameter and chosen as $\tau_{\mathrm{SO}}=2.5\times10^{-13}\,\mathrm{s}$ so that the simulation best reproduces the experimental $B$–$T$ phase boundary; this value remains close to that reported for Sr-doped NdNiO$_2$ in the Supplementary Information of Ref.~[\citen{wang_isotropic_2021}]. Additionally, we use an effective electron mass $m^\ast=3\,m_e$, consistent with Angular resolved photoemission electron spectroscopy (ARPES) data on an infinite layer nickelate films \cite{sun_electronic_2025} and with Density functional theory and Dynamical mean field theory calculations (DFT/DMFT) \cite{kitatani_nickelate_2020}.\\

A crucial input is the total effective field $B_{\rm tot}$ experienced by the Cooper pairs. We define:\\

\begin{equation*}
\begin{aligned}
B_{\rm tot} &= B_0
+ \frac{\Gamma}{g\mu_B}\Bigg(
x_{\rm Eu}\,\frac{(g_{\rm Eu}-1)\,\langle m_{\rm Eu}\rangle}{g_{\rm Eu}\,\mu_B} \\
&\qquad\qquad\;\; + x_{\rm Nd}\,\frac{(g_{\rm Nd}-1)\,\langle m_{\rm Nd}\rangle}{g_{\rm Nd}\,\mu_B}
\Bigg)
\end{aligned}
\end{equation*}
\\

where $\langle m_{Eu}\rangle$ and $\langle m_{Nd}\rangle$ are average magnetization of the two species, $g$ is the Land\'e factor of the Ni ion  and $\Gamma$ is the negative exchange coupling constant, $\Gamma=-1.6\times10^{-21}\,\mathrm{J}$.\\
The obtained $B_{tot}$ embeds both the externally applied field $B_0$ and the \ed{field-induced } exchange contributions from the Eu$^{2+}$ and Nd$^{3+}$ moments. Finally, one inserts $\alpha(B_{\rm tot})=\alpha_1+\alpha_2$ into the gap equation and  solves it numerically to obtain $B_{c2}(T)$ for the perpendicular and parallel field configurations.\\

\begin{figure}[H]
\centering
\includegraphics[width=0.6\textwidth]{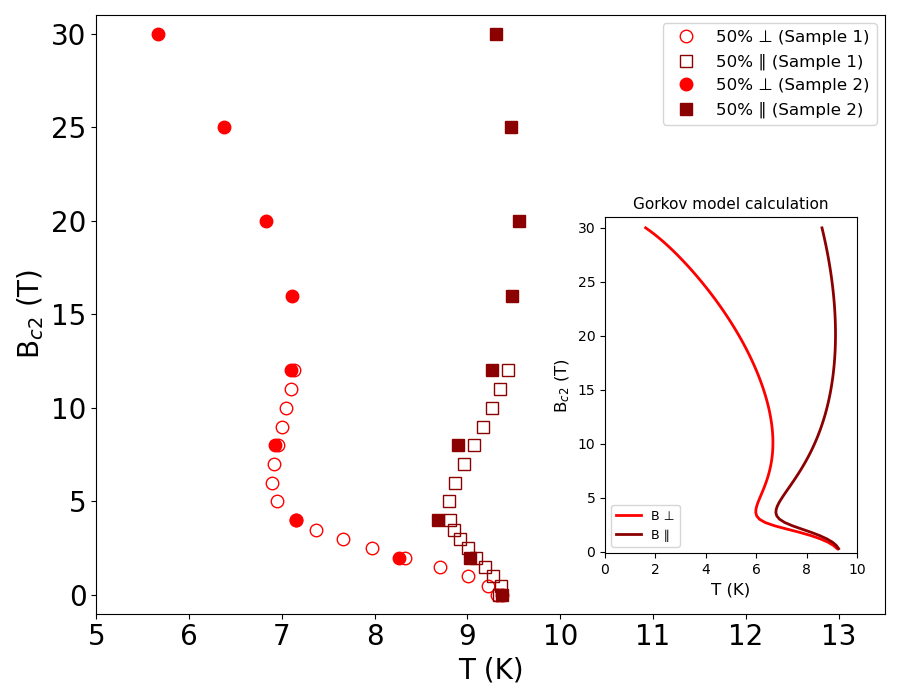}
\caption{Experimental and theoretical parallel and perpendicular critical fields versus temperature for the 50\% resistivity criterion. The experimental data are shown in the main panel while the inset shows the theoretical $B_{c2}(T)$ obtained by solving the modified gap equation. }
\label{fig:5}
\end{figure}

Figure \ref{fig:5} displays for the 50\% resistivity criterion both the experimentally measured  parallel and perpendicular critical fields (main panel) and the calculations described above in the inset. As can be seen, the calculations capture the pronounced non-linearity and low-temperature downturn of $B_{c2}$, thus directly linking the observed re-entrant superconductivity to the competing polarizations of the rare-earth moments.

\end{multicols}

\begin{multicols}{2}

In summary, we have  synthesized and investigated Nd$_{1-x}$Eu$_x$NiO$_2$ thin films grown via RF magnetron sputtering and subsequently reduced \textit{in situ} using a metallic aluminum capping layer to induce the necessary topotactic reaction. This approach allowed high quality  infinite-layer films to be obtained on  LSAT and NdGaO$_3$ substrates.\\
Transport measurements revealed a full superconducting transition in films grown on LSAT. Applying a perpendicular or a parallel magnetic field allowed us to uncover a re-entrant superconducting behavior: \ed{Low field values} (a few T) bring the system into a dissipative phase;  further increase leads to the full recovery of superconductivity.
 In contrast, films grown on NdGaO$_3$ did not show superconductivity down to the base temperature, providing a valuable platform to probe normal-state magnetic and transport properties down to the lowest temperatures investigated.\\ 
 We relate the \ed{unique} superconducting behavior \ed{that we unveiled } to the presence in our samples of two types of magnetic rare-earth ions, Eu$^{2+}$ and Nd$^{3+}$, with opposite influence on the effective field seen by the conduction electrons - a Jaccarino-Peter like effect. 
Quantitative modeling of the (extraordinary) Hall effect (measured on an optimally doped sample grown on NdGaO$_3$) based on the projected spin polarizations of the two rare-earth magnetic species, Eu$^{2+}$ and Nd$^{3+}$, successfully captures the temperature and field dependence of the Hall resistance, indicating that both magnetic subcomponents play an active role in modulating the effective field experienced by the conduction electrons.\

Unlike the original Jaccarino-Peter mechanism in which superconductivity emerges due to the compensation between an external magnetic field and the exchange field of a single magnetic ion specie \cite{jaccarino_ultra-high-field_1962}, our findings suggest a more intricate Jaccarino-Peter-like effect with the compensation arising from the interplay between the spin polarizations of two distinct rare-earth species, leading to a net suppression of the internal field for an intermediate field range. \\
This picture is \ed{further supported } by a microscopic description of the superconducting phase boundary in the $B_{c2}$–$T$ plane using Gorkov’s pair-breaking theory. Considering that the total field \ed{acting on the spin part of the wavefunction} is the sum of the external magnetic field and the exchange field produced by the presence of the two magnetic ions, \ed{allows us to reproduce  the experimentally measured  in-plane and out-of- plane $B_{c2}(T)$ critical field lines} $-$ thereby directly linking the observed magnetic behavior to the unconventional superconducting properties.
We \ed{infer} that the magnetic contributions \ed{of both  Eu$^{2+}$ and Nd$^{3+}$ ions} are central to the observed field-induced re-entrant superconductivity, \ed{underscoring} the complex interplay between local magnetism and superconductivity in Eu-doped infinite-layer nickelates.
\end{multicols}

\paragraph{Methods}
The perovskite nickelates were grown using off-axis magnetron sputtering, as described in reference [\citen{varbaro_electronic_2023}].\\
Metallic aluminum was deposited using an AC-powered sputtering gun. The deposition was performed in a pure argon atmosphere (0.01 Torr) at 270 °C, with a typical growth rate of approximately 0.25 nm/s.
Despite the relatively low growth temperature of the Al layer, compared to other groups, we still observed that the required reduction conditions depended on film thickness: thicker precursor layers \ed{required} higher Al deposition temperatures to achieve complete reduction. Notably, and unlike in earlier findings \cite{zhang_achieving_2024}, we found that increasing the Al deposition rate significantly improved film quality. This was achieved by increasing the RF power of the Al gun from 100 W to 150 W, allowing the deposition of a nominal Al thickness of $\sim$3–3.5 nm in approximately 14 seconds. All reductions were performed without any post-annealing, again in contrast with several literature reports involving prolonged thermal treatments \cite{zhang_achieving_2024, li_superconducting_2020, wei_solid_2023, zeng_phase_2020}. We \ed{tentatively} attribute this difference to the relatively high base pressure of our sputtering chamber ($\sim$5 × 10$^{-6}$ mbar), which may affect the reduction kinetics.\\XAS and XMCD measurements were performed at the BOREAS beamline at the ALBA synchrotron in Barcelona, Spain [\citen{barla_design_2016}]. The experiments were carried out in grazing incidence geometry (70° to the sample normal) with 90\% circularly left- and right- polarized light. The XAS data are defined as the average of the absorption spectra obtained from the two circular polarizations. The XMCD signal is defined as the difference between the absorption spectra of the two polarizations, normalized by the maximum XAS signal.

\paragraph{Acknowledgments:}
We would like to thank Giacomo Sala and Stefano Gariglio for the their insights and enlightening discussions.\\ We also thank Marco Lopes for very efficient technical support.\\ 
This work was supported by the Swiss National Science Foundation – division II (200020\_179155 and 200020\_207338).
The authors acknowledge access to the electron microscopy facilities at the Interdisciplinary Centre for Electron Microscopy, École Polytechnique Fédérale de Lausanne.

\bibliography{Thesis3}
\bibliographystyle{unsrt}

\end{document}